\documentclass[aps,preprint,superscriptaddress]{revtex4-1} 

\usepackage{graphicx} 
\usepackage{mathtools}
\usepackage{mathptmx}
   \usepackage{lineno}
\usepackage{color}
\begin{document}

\title{Driven dust vortex characteristics in plasma with 
external transverse and weak magnetic field}

\author{Modhuchandra Laishram}\email{modhu@ipr.res.in}
\address{Institute for Plasma Research, HBNI, Bhat, Gandhinagar, India, 382428}
\date{\today}
\begin{abstract}
The two-dimensional hydrodynamic model for bounded 
dust flow dynamics in plasma is extended for analysis of driven vortex characteristics in presence of external transverse and weak magnetic 
field (${\bf B}$) in a planner setup and parametric regimes motivated by recent magnetized dusty plasma (MDP) experiments.
This analysis has shown that shear in the ${\bf B}$ can produce a shear internal field (${\bf E_a}$) in between electrons and ions due to the ${\bf E}\times {\bf B}$ and ${\nabla {\bf B}}\times {\bf B}$-drifts that cause rotation of dust cloud levitated in the plasma. The flow solution demonstrates that neutral pressure decides the dominance between the ions-drag and the ${\bf E_a}$-force. The shear ions-drag generates an anti-clockwise circular vortical structure, whereas the shear ${\bf E_a}$-force is very localized and gives rise to a clockwise $D$-shaped elliptical structure which turns into a meridional structure with decreasing ${\bf B}$. Effect of system parameters including the strength of ${\bf B}$ by varying its magnitude and shear mode numbers and the sheath field are analyzed within the weak MDP regime, showing noticeable changes in the flow structure and its momentum. In the parametric regime of high pressure and lower ${\bf B}$, the ${\bf E_a}$-force becomes comparable or dominant over the ion drag and peculiar counter-rotating vortex pairs are developed in the domain. Further, when the ${\bf B}$ is flipped by $180^0$-degree, both the drivers act together and give rise to a single strong meridional structure, showing the importance of ${\bf B}$-direction in MDP systems. It further discussed similar elliptical/meridional structures reported in several MDP experiments and relevant natural driven-dissipative flow systems.
\end{abstract}

\maketitle
\section{Introduction}
\label{introduction}
Dusty plasma which is partially ionized and low-temperature plasma consists of additional suspended micron-sized charged dust particles 
are known for being an ideal domain for experimental as well as theoretical studies of basic collective behavior in relevant nonequilibrium flows system of 
nature~\cite{RevModPhys.81.1353,shukla2015introduction,Saitou2020,
Nivedita@Dean_flow2017,PhysRevE.81.041124,PhysRevE.91.063110,PhysRevE.95.033204}.  
When the suspended dust particles are numerous, the interaction with the background plasma and neutral gas constitutes a range of interesting 
fluid-phases phenomena such as waves, instabilities, vortices, void, and other nonlinear structures~\cite{shukla2015introduction,PhysRevLett.83.1598,PhysRevLett.111.185003, PhysRevE.59.1047,doi:10.1063/1.5042497}. In particular, the vortices are ubiquitous in nature from small scale biological complexity to large scale planetary surface~\cite{Nivedita@Dean_flow2017,doi:10.1146/annurev-fluid-122414-034558,marcus_1990,PhysRevE.95.033204,Laishram_2019}. The 
vortex structure in the nonequilibrium systems evolves subjected to various driving fields and the dynamical regimes as in the dusty plasma~\cite{Nivedita@Dean_flow2017,PhysRevE.81.041124,doi:10.1063/5.0004842}. The vortical structure has been observed in different shapes, sizes, numbers, orientations, and strength~\cite{doi:10.1146/annurev-fluid-122414-034558,doi:10.1063/5.0004842,PhysRevLett.101.235001}. For examples, the elliptical flow structure are observed in magnetized dusty plasma (MDP) systems~\cite{PhysRevLett.111.185003,doi:10.1063/5.0004842}, which is also common in magnetic confinement configurations in a tokamak, FRC (field reverse configuration)~\cite{8692190,doi:10.1063/1.3613680}, and in neutral fluid flows such as the sedimentation of dust in a lid-driven cavity at low to moderate Reynolds number~\cite{roy2020improved,articleMat,PhysRevE.81.041124}. Further, the vortex structure plays an important role in the fluid mixing and transport process in laminar and turbulent flows~\cite{Nivedita@Dean_flow2017,doi:10.1146/annurev-fluid-122414-034558}. Therefore, it has been a topic of active research that how the vortical structure evolves with the system parameters in various complex fluid flow. Using the dusty plasma, numbers of our past studies have investigated the characteristic features of the vortex in nonequilibrium systems~\cite{PhysRevE.91.063110,PhysRevE.95.033204,Laishram_2019}. However, the underlying physics of elliptical vortical structures are yet to be fully understood.

Recently, the elliptical vortical structures are reported in recent MDP experiments~\cite{doi:10.1063/5.0004842,PhysRevLett.111.185003}, and there is a resurgent of interest in the studies of dust collective behaviors in the MDP because of its importance and varieties of applications in studies of astrophysical star formation, space, laboratory discharge, fusion devices, and industrial processing~\cite{merlino2006dusty,Cohen_2012,Krasheninnikov_2011,doi:10.1063/1.1587887}. For examples, Mark Kushner~\cite{doi:10.1063/1.1587887} and Barnat {\it et~al.}~~\cite{Barnat_2008} analyzed the RF discharge of the industrial etching process under the influence of a transverse magnetic field, showing the plasma becomes more resistive to electrons and the distribution becomes skewed toward the $\bf{(E\times B)}$-drift. The mobility of electrons and ions are different, therefore, charge separation is created along the $\bf{(E\times B)}$-drift in presence of the collisions. 
Thus, a new internal electric field called ambipolar electric field ${\bf E_a}$ is arisen in between the magnetized electrons and the ions. 
In space science, atmospheric escape remains an open issue, and Cohen~\cite{Cohen_2012} reported the transport of ionospheric constituents to Earth’s (as well as other in other planets) magnetosphere by the similar ambipolar ${\bf E_a}$ generated in the system. Therefore, a detailed analysis of the ${\bf E_a}$ and Lorentz force acting on the dust particles other than the usual ions and neutral drags is crucially important for understanding the relevant physical process in nature.  

One main difficulty in the MDP is that the domain is widely diverse starting from strongly magnetized to weakly magnetized systems, from dominant longitudinal to transverse magnetic fields, and many more. In most of the past MDP-studies where dust vortices are experimentally observed, the magnetic field is considered parallel to the sheath electric field (${\bf E_s}$) i.e., ${\bf B}\|{\bf E_s}$ and both the fields are directed along with the gravity~\cite{sato2001dynamics,PhysRevE.61.1890,PhysRevLett.111.185003,doi:10.1063/5.0004842}. In these experiments, dust is not magnetized, however, 
the $\bf{(E_s\times B)}$-drift of the magnetized ions excited the 
rotational motion of the dust particles. In recent years, a lot of 
efforts has been made to fully magnetize heavy dust in MDP 
laboratories~\cite{Thomas_2012,Thomas_2019,doi:10.1063/5.0004842}, although it is big 
challenges as the dust is relatively heavy and so require a strong 
magnetic field (B $\ge$ 4 Testla)~\cite{Thomas_2012}. On the other, there 
has been less discussion in MDP with the transverse magnetic field where the ${\bf B}\perp {\bf E_s}$. Yeng and Maemura {\it et~al.}~\cite{Yang_1996,MAEMURA19981351} discussed the transport of 
the dust particles raised by the $\bf{(E_s\times B)}$-drift of electrons in the system. Samsonov {\it et~al.}~\cite{Samsonov_2003} analyzed the enhancement of dust levitation of by an inhomogeneous 
magnetic field$(\bf \nabla B)$. Also, the inhomogeneity of density $(\nabla n)$ is 
common in realistic low-temperature plasma with dust particles~\cite{Losseva_2008}. Further, in the experiment by Puttscher and Melzer {\it et~al.}~\cite{Puttscher_2014}, 
the dynamics of dust particle displacement are investigated in presence of the transverse and weak ${\bf B}$ i.e., only electrons are magnetized. They observed the displacement of single particles and whole dust clusters by the ${\bf \nabla B}$-drift and ambipolar $\bf{(E\times B)}$-drift due to the magnetization of the electrons.  Further, In their subsequent work~\cite{doi:10.1063/1.4981928,doi:10.1063/1.4904039}, they discussed 
the competition between the ambipolar $\bf{(E\times B)}$-drift and ion/neutral drag using a force balance model that decides the displaced direction of the dust cluster. However, there are many queries if 2D vortex of the dust 
cloud can be developed in the same configuration? What will be the effect of diamagnetic-drift and the gradient-drift which are very common in realistic laboratory plasma~\cite{chen2015introduction}? Most importantly what will be the characteristics of dust vortex structure in the setup with transverse and weak ${\bf B}$? To know 
the physics insight of the above queries, systematic studies using a theoretical and numerical analysis of the MDP system is required as attempted in 
the present work.

\begin{figure}
\includegraphics[width=.69\textwidth]{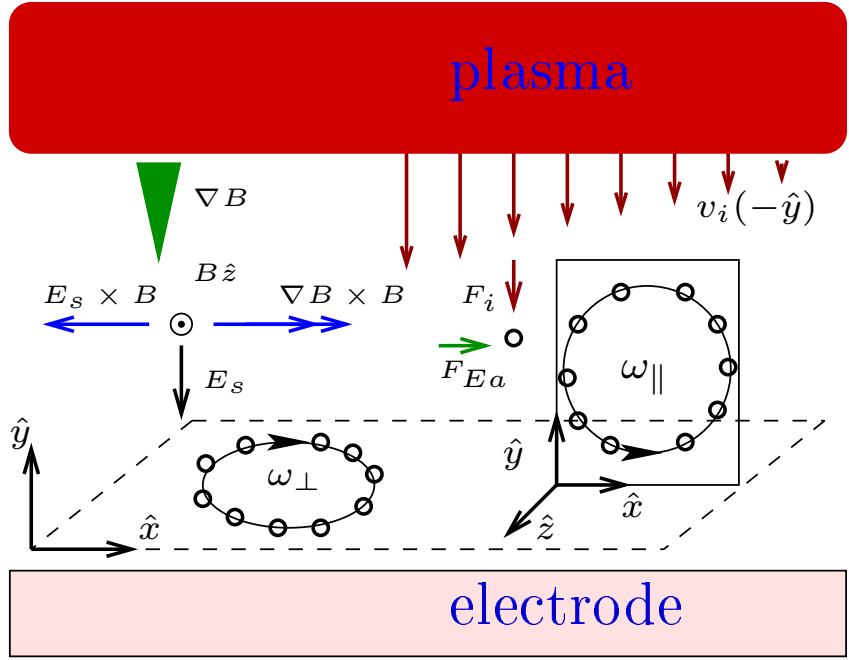}
\caption{Schematic representation of dust cloud levitated above the electrode  
by an electrostatic potential $\phi_b$ in presence of sheath 
field ${\bf E_s~(-\hat y)}$, transverse 
magnetic field ${\bf B~\hat z}$ having shear ${\bf \nabla B~ \hat y}$. The main driving fields are 
ion-drag force ${\bf F_i~(-\hat y)}$ and the internal ambipolar field ${\bf F_{Ea}~\hat x}$ throughout the dust domain. 
 \label{schematic}}
\end{figure}

Studies of steady-state dust vortex characteristics in a non-magnetized 
and incompressible fluid-phase regime of dusty plasmas have been 
extensively done in a series of our theoretical-simulation 
work~\cite{Laishram_2019,PhysRevE.95.033204,PhysRevE.91.063110,doi:10.1063/1.4887003}. In the current work, we extend the existing model in presence of external transverse and weak ${\bf B}$ in a planner setup and parametric regimes motivated by recent magnetized dusty plasma(MDP) experiments~\cite{doi:10.1063/1.4981928,doi:10.1063/1.4904039,Puttscher_2014,sato2001dynamics,PhysRevE.61.1890,doi:10.1002/ctpp.200910017}. This formulation reveals the conditions for 
sustaining a 2D vortex of the dust cloud in a cross-section parallel ($\omega_{\|}$) and 
perpendicular ($\omega_{\perp}$) to the ${\bf B}$ as highlighted in Fig.~\ref{schematic}. Shear in the ${\bf B}$ can produce a shear internal field ${\bf E_a}$ in between electrons and ions due to the combined ${\bf E_s}\times {\bf B}$ and ${\nabla {\bf B}}\times {\bf B}$-drifts that causes a rotation of the dust cloud levitated in the plasma. 
Effect of system parameters including the strength of shear ${\bf B}$ and the ${\bf E_s}$ are analyzed within the weak ${\bf B}$ regime ($\omega_{ec}\ge\nu_{en}$ to $\nu_{in}\ge\omega_{ic} $, where, $\omega_{jc}$ is the cyclotron frequency and $\nu_{jn}$ is collision frequency of $j$-species with neutrals), demonstrating the role of ${\bf B}$-direction and formation of peculiar counter-rotating vortex pairs developed in the same setup.

This paper is organized as follows. In Sec.~\ref{formulation}, 
we discuss the extended 2D hydrodynamic model for the 
bounded dust flow in a MDP setup, deriving the general equation of 
the shear field ${\bf E_a}$ and associated vorticity sources. The 2D dust vortex solutions are characterized concerning specific driving fields and parametric regimes in Sec.~\ref{Simul_result}. The circular 
anti-clockwise structure due to the dominant ion-drag force is analyzed in Sec.~\ref{Results_1}, whereas the elliptical/meridional structure due to a dominant shear ${\bf E_a}$-force is discussed in Sec.~\ref{structure_all_B}. Further, a new condition for the formation of counter-rotating vortex pairs and the impact of the ${\bf E_s}$ are discussed in Sec.~\ref{structure_all_B_i}. 
Summary and conclusions are presented in Sec.~\ref{conclusion}.
\section{HYDRODYNAMICAL MODEL OF DUST DYNAMICS IN A MAGNETIZED PLASMA}
\label{formulation}
As discussed above, the present work is motivated by several new dusty plasma experiments~\cite{Yang_1996,MAEMURA19981351,Puttscher_2014,doi:10.1063/1.4981928,doi:10.1063/1.4904039} which have studied the behavior of dust particles dynamics applying transverse ${\bf B}$ near the sheath region of a glow discharge plasma. Therefore, we consider a similar system of dust cloud/cluster suspended near the sheath region of magnetized plasma. The schematic cross-section of the system is shown in Fig.~\ref{schematic}, which has demonstrated that dust particles are trapped near the sheath region 
using an electrostatic potential $\phi$ and a transverse magnetic 
field ${\bf B}\hat z$ perpendicular to the sheath 
field ${\bf E_s}(-\hat {y})$ is applied to the system. The detailed description of the dusty plasma experimental system with the transverse ${\bf B}$ is given in the Ref.~\cite{doi:10.1063/1.4981928} and Ref.~\cite{doi:10.1063/1.4904039}.

Now, we extend the hydrodynamic model for the dust flow from our 
previous work considering the weak transverse ${\bf B}$ ~\cite{PhysRevE.95.033204}. For a bounded dust cloud that 
satisfies 
incompressible, 
isothermal conditions, 
and has a finite viscosity, the dynamics 
in the magnetized plasma can be model by the simplified continuity 
equation and Navier-Stokes momentum equation as follows~\cite{landau2013fluid,PhysRevE.95.033204,doi:10.1063/1.5045772},
\begin{eqnarray}
\nabla\cdot{\bf u}=0, 
\label{cont-equation}
\end{eqnarray}
\begin{eqnarray}\nonumber
\frac{\partial {\bf u}}{\partial t}  
+ ({\bf u} \cdot \nabla) {\bf u} 
=-\nabla{\phi_b} +\frac{q_d}{m_d}{\bf E_a} +\frac{1}{\rho}({\bf J_d}\times {\bf B}) - \frac{\nabla{P}}{\rho}\\  
+ \mu \nabla^2 {\bf u}  
 - \xi ({\bf u} - {\bf v})  - \nu ({\bf u} - {\bf w}).
\label{ns-equation}
\end{eqnarray}

Here ${\bf u}$, ${\bf v}$, and ${\bf w}$ are the flow velocities of the dust, ion, and neutral fluids, respectively. ${\phi_b}$ is the effective confining potential (from all the conservative fields including gravity). $q_d$ is dust charge, $m_d$ is dust mass, and ${\bf J_d}=n_dq_d{\bf u}$ is dust current density. ${\bf E_a}$ is the internal field due to charge separation 
between electrons and ions of the background plasma due to 
the magnetization~\cite{Losseva_2008}. 
$P$ and $\rho$ are the
pressure and mass density of the dust fluid, respectively, $\mu$ is the kinematic viscosity, $\xi$ is the coefficient of ion-drag acting on the dust, and $\nu$ is the coefficient of friction generated by the background neutral fluid
\cite{PhysRevLett.68.313,PhysRevE.66.046414,PhysRevLett.92.205007}.
The above Eq.~(\ref{ns-equation}) describes that the dust cloud can be confined near the sheath region of the discharge by balancing all the conservative fields $\phi_b$, and it can be driven either by the 
${\bf E_a}$-force  or by the interaction with the background ion or neutral drag present in the system. The proposed model consists of various forces term is motivated by recent works in magnetized dusty plasma~\cite{PhysRevLett.111.185003}.
We know that during the longer time scale on which the steady dust flow is maintained, all the other highly mobile fluids such as electrons and ions have already been in thermal equilibrium even though it interacts with the confined dust fluid and drive it. Therefore, the steady-state profile of ${\bf E_a}$ of electron magnetization and ions velocity field ${\bf v}$ in Eq.~(\ref{ns-equation}) is valid for the analysis of steady dust flow characteristics in the weakly magnetized 
dusty plasma.  

\subsection{Vorticity-Stream function formulation}
\label{Vorticity_model}
Our main goal is to study the characteristics of 2D vortex or circulating motion of a dust cloud/cluster confined in a weak magnetized dusty plasma. 
Therefore, taking curl of the the above equation~(\ref{ns-equation}), the steady dust flow in a 2D plane either in ${\bf xy}$ or ${\bf xz}$ can be written in term of dust stream function ($\psi$) and corresponding vorticity ($\omega$) as follows~\cite{PhysRevE.91.063110,PhysRevE.95.033204},
\begin{eqnarray} 
\nabla^2\psi&=&-\omega,
\label{streamf-equation}\\
\frac{\partial {\bf \omega}}{\partial t} 
+({\bf u} \cdot \nabla) {\omega}&=&\mu \nabla^2 \omega - 
(\xi+\nu) \omega +\xi \omega_s +\beta\omega_B. 
\label{vort-equation}
\end{eqnarray}
%
Where, ${\bf u}=\nabla\times \psi$, $\omega=\nabla\times {\bf u}$, and
the direction of the $\psi$ and $\omega$ is always perpendicular to the 
chosen 2D plane. The curl of terms having conservative fields such as $P$ and $\phi$ goes to zero. It means
conservative forces do not contribute to any vortex dynamics and its other roles are absorbed in the streamfunction and vorticity fields. $\omega_{s}$ is the external vorticity source from the 
curl of $\xi{\bf v}$ and $\nu{\bf w}$ terms, i.e., from the shear nature of unbounded background flow fields. For simplicity, the neutral velocity field ${\bf w}$ is assumed to be stationary or we are on the neutral frame of references. In a general electrostatic fluid constituted by charged particles, the coefficients $\xi$ and $\nu$ depend on the state variables such as 
temperature, densities, velocity, and charge distribution
\cite{PhysRevLett.68.313,PhysRevE.66.046414,
PhysRevLett.92.205007}. 
As a consequence, the {$\omega_s$} in the 
Eq.~(\ref{vort-equation}) represents any non-zero fields, 
such as $\nabla\times {\bf u}_{i(n)}$,
$\nabla Q_d\times{\bf E}$, 
$\nabla u_{i(n)} \times \nabla n_{i(n)}$, 
or their effective combinations as the external driving mechanism for the dust vorticity in a non-magnetized dusty plasma~\cite{Laishram_2019}. Where $Q_d$ is dust charge, {${\bf E}$} is an effective electric field along with the streaming ions with velocity ${\bf u_i}$ and density ${n_i}$. 
The subscript $i(n)$ represents the background ions (neutrals). Further, $\omega_{B}$ is the additional vorticity source from the curl of magnetization forces, and $\beta$ is the co-efficient of magnetization. The detail derivation of $\beta\omega_{B}$ is discussed in the following section~\ref{Vorticity_Ws}.

\subsection{Vorticity sources ($\beta\omega_{B}$) in a weakly magnetized plasma}
\label{Vorticity_Ws}
From the above equations (\ref{ns-equation}) and (\ref{vort-equation}), 
the addition vorticity source $\beta\omega_{B}$ can be written as follow,
\begin{eqnarray}
 \beta\omega_{B}=\nabla\times\left[\frac{q_d}{m_d}{\bf E_a} 
 +\frac{1}{\rho}({\bf J_d}\times {\bf B})\right]
 \label{beta_omegaB}
\end{eqnarray}
In the weakly magnetized dusty plasma, the last term, i.e., Lorentz forces on dust particle is neglected for the rest of the present analysis. 
Then, for the two-dimensional plasma flow 
across ${\bf B}$, it is known that charge separation usually takes place due to available drifts (such as ${\bf E \times B}$, ${ \nabla n_j \bf\times B}$ , and ${\bf \nabla B \times B}$ present in the system)~\cite{chen2015introduction}, and a charge separation field (${\bf E_a}$) is arised in the system. To estimate the ${\bf E_a}$, we start from the perpendicular component of flow equation of motion of electrons and ions as follows,
\begin{eqnarray} \nonumber
 m_jn_j\frac{d{\bf u_{j\perp}}}{dt} =
q_jm_j({\bf E_\perp} +{\bf u_{j\perp}}\times {\bf B})- K_bT_j{\nabla{n_j}} + m_jn_j\nu_{jn}{\bf u_{j\perp}}  
\label{es-equation}
\end{eqnarray}
Where the subscript $j$ represents the background electrons and ions with mass $m_j$, density $n_j$, and drift velocities $ {\bf u_{j\perp}}$ across the ${\bf B}$. 
Assuming isothermal and weakly ionization plasma, the corresponding  $u_{j\perp}$ of the $jth$ species 
can be derived~\cite{chen2015introduction} as,
\begin{eqnarray}
 u_{j\perp} =\pm\mu_{j\perp}{\bf E_\perp} 
 -D_{j\perp}\frac{\nabla{n_j}}{n_j} 
 +\frac{{\bf v}_{jB}+{\bf v}_{jD}+{\bf v}_{j\nabla B}}
 {1+(\nu_{jn}^2/\omega_{jc}^2)}.  
\label{es-equation2}
\end{eqnarray}
Here, $\mu_{j\perp}=\frac{\mu_j}{1+\omega_{jc}^2\tau_{jn}^2},~ 
\mu_j= q_j/m_j\nu_{jn}$, \\
$D_{j\perp}=\frac{D_j}{1+\omega_{jc}^2\tau_{jn}^2},~ D_j= K_bT_j/m_j\nu_{jn}$, \\
${\bf v}_{jB}= \frac{{\bf E}\times{\bf B}}{|{\bf B}|^2}$, 
${\bf v}_{jD}= \mp\frac{K_bT_j}{|q_j|{|{\bf B}|^2}}
\frac{\nabla n_j\times {\bf B}}{n_j}$, 
${\bf v}_{\nabla B}=\mp\frac{r_{j\small L}{v}_{j\small L}}{2} \frac{\nabla {\bf B}\times {\bf B}}{{|\bf B}|^2}$,\\ 
$r_{j\small L}=m_j{v}_{j\small L}/|q_j|B $, and 
$m_j{v}_{j\small L}^2/2\approx K_bT_j$.
Where,  $\mu_{j\perp}$ and $D_{j\perp}$ are mobility and diffusivity across the magnetic field respectively. $\omega_{jc}=q_jB/m_j$ is Larmour frequencies and $\tau_{jn}=1/\nu_{jn}$ is the collision time of $jth$ species with background neutrals. ${\bf v}_{jB}$ is the (${\bf E}\times{\bf B}$)-drift across the plane containing ${\bf E}$ and ${\bf B}$, ${\bf v}_{jD}$ is the diamagnetic-drift across the magnetic field, and ${\bf v}_{j\nabla B}$ is the gradient-drift of the magnetic field. Both ${\bf v}_{jD}$ and ${\bf v}_{j\nabla B}$ are charge dependence, whereas ${\bf v}_{jB}$ is independent of charge. In the 
Eq.~\ref{es-equation2}, the first term and second term on the right side are immediately recognizable as drift along the gradients in potential and density which are reduced by the factor ($1+\omega_{jc}^2\tau_{jn}^2$) due to the magnetization. The third term consisting of ${\bf v}_{jB}$, ${\bf v}_{jD}$, and ${\bf v}_{j\nabla B}$ are drift across the gradients in potential and density which are reduced by a different factor ($1+1/\omega_{jc}^2\tau_{jn}^2$), and all the contributions are slowed down by collisions with neutrals and magnetization.
Ambipolar diffusion is not a trivial problem in presence of inhomogeneous ${\nabla n}$ and ${\bf \nabla B}$. In highly magnetized 
plasma~\cite{doi:10.1063/5.0004842,PhysRevLett.111.185003}, it would be expected that ions move faster than electrons across the magnetic field. However, in the weak magnetization, electrons still move faster than ions due to their mobility difference and collision with neutrals. Therefore, for the present 2D flow analysis in presence of weak magnetic field, 
we consider the condition for ambipolar diffusion 
and quasi-neutrality condition 
in a dusty plasma. Then the corresponding expression for ambipolar field ${\bf E_a}$ can be 
obtained as follow,
 \begin{eqnarray}\nonumber
 {\bf E_a} = \eta q_e(D_{i\perp}{\nabla{n_i}}-D_{e\perp}{\nabla{n_e}}) \\   
 +\frac{\eta q_e}{(1+\nu_{en}^2/\omega_{ec}^2)}\bigg[n_e\frac{({\bf E}\times{\bf B})}{|{\bf B}|^2} 
  +\frac{K_bT_e}{q_e}
 \frac{(\nabla n_e\times {\bf B})}{{|{\bf B}|^2}}
 +\frac{n_e K_bT_e}{q_e} 
 \frac{\nabla {\bf B}\times {\bf B}}{{|\bf B}|^3}\bigg].
 \label{E_a_eqn}
\end{eqnarray}
Where $ q_e\eta= 1/(n_i\mu_i+n_e\mu_e$), $\eta$ is the plasma resistivity. In the weakly magnetized plasma, $1+(\nu_{en}^2/\omega_{ec}^2)$ is an important quantity. 
It shows the influence of magnetization depend on the ratio of the collision and cyclotron frequencies. 
Taking curl, the above equation can be expressed as, 
\begin{eqnarray}\nonumber
  \nabla\times{\bf E_a} = \frac{\eta q_e}{(1+\nu_{en}^2/\omega_{ec}^2)} \bigg[ n_e\nabla\times\frac{({\bf E}\times{\bf B})}{|{\bf B}|^2} 
  +\frac{K_bT_e}{q_e}
 \nabla\times\frac{(\nabla n_e\times {\bf B})}{{|{\bf B}|^2}} 
 +\frac{n_e K_bT_e}{q_e} 
\nabla\times \frac{(\nabla {\bf B}\times {\bf B})}{{|\bf B}|^3}
\bigg].
\label{cross_Ea_eqn}
\end{eqnarray}

Now, using the standard vector identity, 
\begin{eqnarray}\nonumber
 \nabla\times ({\bf A}\times{\bf B}) = {\bf A}(\nabla\cdot{\bf B})
  -{\bf B}(\nabla\cdot{\bf A})+({\bf B}\cdot\nabla){\bf A} 
  - ({\bf A}\cdot\nabla){\bf B},
\end{eqnarray} 
the $1st$-term of the Eq.~(\ref{cross_Ea_eqn}) can be expressed as,
{\small
\begin{eqnarray} 
 \nabla\times\frac{({\bf E}\times{\bf B})}{|{\bf B}|^2} 
 &= {\bf E}(\nabla\cdot{\bf\frac{B}{|B|^2}})
  -{\bf\frac{B}{|B|^2}}(\nabla\cdot{\bf E})+({\bf\frac{B}{|B|^2}}\cdot\nabla){\bf E} 
  - ({\bf E}\cdot\nabla){\bf\frac{B}{|B|^2}}.~~~~~  
\label{cross_EB}
\end{eqnarray}
}
It is to be noted that, the first term on the right-hand side is negligible as the ${\bf B}$ is divergenceless. Similarly, the $2nd$-term and $3th$-term of the Eq.~(\ref{cross_Ea_eqn}) can be expressed as follows,
{\small
\begin{eqnarray}
 \nabla\times\frac{({\nabla n_e}\times{\bf B})}{|{\bf B}|^2} 
  =  -{\bf\frac{B}{|B|^2}}(\nabla^2 n_e)
+({\bf\frac{B}{|B|^2}}\cdot\nabla){\nabla n_e}
- ({\nabla n_e}\cdot\nabla){\bf\frac{B}{|B|^2}}. ~~~~~
  \label{cross_nB}
\end{eqnarray}
\begin{eqnarray}
 \nabla\times\frac{({\nabla \bf B}\times{\bf B})}{|{\bf B}|^3} 
  =  -{\bf\frac{B}{|B|^3}}(\nabla^2 \bf B)
+({\bf\frac{B}{|B|^3}}\cdot\nabla){\nabla {\bf B}}
  - ({\nabla \bf B}\cdot\nabla){\bf\frac{B}{|B|^3}}.~~~~
  \label{cross_bB}
\end{eqnarray}
}
Putting Eqs.~(\ref{cross_EB}), ~(\ref{cross_nB}), and ~(\ref{cross_bB}) in the Eq.~(\ref{cross_Ea_eqn}), further from Eq.~(\ref{cross_Ea_eqn}) and Eq.~(\ref{beta_omegaB}), we can write for the vorticity  source as, 
%
%
 \begin{eqnarray}\nonumber
 \beta\omega_{B} = \frac{\eta q_eq_d}{m_d(1+\nu_{en}^2/\omega_{ec}^2)}
\bigg[-n_e\bigg({\bf\frac{B}{|B|^2}}(\nabla\cdot{\bf E})
-({\bf\frac{B}{|B|^2}}\cdot\nabla){{\bf E}} 
+ ({\bf E}\cdot\nabla){\bf\frac{B}{|B|^2}}\bigg)\\ \nonumber
-\frac{K_bT_e}{q_e}\bigg({\bf\frac{B}{|B|^2}}({\nabla^2 n_{e}})
-({\bf\frac{B}{|B|^2}}\cdot\nabla){\nabla { n_e}}
+({\nabla n_e}\cdot\nabla){\bf\frac{B}{|B|^2}}\bigg)\\
 -\frac{n_eK_bT_e}{q_e} \bigg(
  {\bf\frac{B}{|B|^3}}({\nabla^2 {\bf B}})
  -({\bf\frac{B}{|B|^3}}\cdot\nabla){\nabla {\bf B}}
 +({\nabla \bf B}\cdot\nabla){\bf\frac{B}{|B|^3}}\bigg)
\bigg].~~~~~~~
\label{cross_Ea_eqn3}
\end{eqnarray}

In a more specific way, we focus on the effect of the weak magnetization on the ${\bf xy}$-plane having the ${\bf E_s}{(-\hat y)}$, ${\bf \nabla B}{\hat y}$ across the transverse ${\bf B}{\hat z}$, then the above Eq.~(\ref{cross_Ea_eqn3}) is reduced as,
{
 \begin{eqnarray}\nonumber
 \beta\omega_{B\|} = \frac{\eta q_eq_d}{m_d(1+\nu_{en}^2/\omega_{ec}^2)} \bigg[-n_e\bigg({\bf\frac{B}{|B|^2}}\frac{\partial{\bf E_s}}{\partial y}
+{\bf E_s}\frac{\partial}{\partial y}{\bf\frac{B}{|B|^2}}\bigg)\\ \nonumber
-\frac{K_bT_e}{q_e}\bigg({\bf\frac{B}{|B|^2}}\frac{\partial^2 n_e}{\partial y^2}+ {\frac{\partial n_e}{\partial y}}\frac{\partial}{\partial y}{\bf\frac{B}{|B|^2}} \bigg)\\
-\frac{n_eK_bT_e}{q_e}\bigg({\bf\frac{B}{|B|^3}}\frac{\partial^2 {B}}{\partial y^2}+ {\frac{\partial {\bf B}}{\partial y}}\frac{\partial}{\partial y}{\bf\frac{B}{|B|^3}} \bigg)
\bigg].~~~~
\label{cross_Ea_eqn4}
\end{eqnarray}
}

 In the above Eq.~(\ref{cross_Ea_eqn4}), the $1st$ term represents the contribution from the variation of the sheath field ${\bf E_s}$ and the $2nd$-term represents the contribution from shear nature of the ${\bf B}$ along the ${\bf E_s}$. The $\partial^2/\partial y^2$ in the $3th$ and the $5th$-terms represent the measure of the concavity of the density/magnetic fields across the field. The $4th$ and the $6th$-terms represent the contribution from shear nature of ${\bf B}$ along with 
 the shear $\partial n/\partial y$ and $\partial {\bf B}/\partial y$ respectively. 
 
 Similarly, when we focus on the ${\bf xz}$-plane along the tranverse ${\bf B}$, i.e., the cross-section parallel to the electrode surface, the above Eq.~(\ref{cross_Ea_eqn3}) can be simplified as follow,

{
 \begin{eqnarray}\nonumber
 \beta\omega_{B_\perp} = \frac{\eta q_eq_d}{m_d(1+\nu_{en}^2/\omega_{ec}^2)} \bigg[
n_e\bigg({\bf\frac{B}{|B|^2}}\frac{\partial}{\partial z}\bigg){\bf E_s}\\
+\frac{K_bT_e}{q_e}\bigg({\bf\frac{B}{|B|^2}}\frac{\partial}{\partial z}\bigg)\frac{\partial n_e}{\partial y}
+\frac{n_eK_bT_e}{q_e}\bigg({\bf\frac{B}{|B|^3}}\frac{\partial}{\partial z}\bigg)\frac{\partial{\bf B}}{\partial y}
\bigg].~~~~~~~~
\label{cross_Ea_eqn5}
\end{eqnarray}
}
In the above Eq.~(\ref{cross_Ea_eqn5}), the $1st$ term represents the 
contribution from ${\bf E_s}$ shear along the ${\bf B}$. 
The $2nd$ and the $3th$-terms represent the contribution from shear 
nature of $\partial n_e/\partial y$ and $\partial {\bf B}/\partial y$ 
along the ${\bf B}$ respectively. Both the terms are contributed only 
when diamagnetic and gradient drift is significant. Again, there 
is no contribution of shear $\partial n_e/\partial x$ and 
$\partial {\bf B}/\partial x$ on the plane. Therefore, in 
the ${\bf xz}$-plane with no finite variation 
of ${\bf E_s}$, ${\partial n_e}/{\partial y}$, 
and ${\partial {\bf B}}/{\partial y}$ along the $\hat z$-direction, 
the 2D dust vortex driven by the 
shear ${\bf E_a}$ is not possible i.e., $\omega_{B_\perp}\approx 0$. 
This may be a good reason 
why dust rotation was not observed in the MDP experiments by Puttscher and Melzer {\it et~al.}~\cite{doi:10.1063/1.4981928,doi:10.1063/1.4904039}.
\section{Dust vortex characteristics in the ${\bf xy}$-plane 
across the transverse ${\bf B}$}
\label{Simul_result}
We are now in the position to study several characteristic features of dust 
vortex in the weakly magnetized plasma. The 2D dust vortex 
dynamics in the ${\bf xy}$-plane across the ${\bf B}$ can be obtained by solving Eqs.~(\ref{streamf-equation}) and (\ref{vort-equation}) using an appropriate profile of the sources fields $\omega_s$, $\omega_{B\|}$, 
boundary conditions, and parameters regime from the 
relevant experimental systems. In the first instance, we define 
specific driving fields and appropriate system parameters following the recent dusty plasma laboratory experiments~\cite{doi:10.1063/1.4981928,doi:10.1063/1.4904039,PhysRevLett.68.313,
PhysRevLett.92.205007,doi:10.1063/1.5078866}.
\subsection{Driving fields and System parameters}
\label{Driver_BC_parm}
 Assuming that the bounded dust cloud/cluster is driven by both the driving 
 fields, i.e., $\omega_s$ from the shear nature of the background 
 ion-drag and $\omega_{B\|}$ from the shear ${\bf E_a}$ derived in Eq.~(\ref{cross_Ea_eqn4}). As in the previous work~\cite{Laishram_2019}, ions are considered streaming downward through the dust cloud having a monotonic shear profile given by a single natural mode of the cartesian plane as given by,
{
\begin{eqnarray}
{\bf v}_{i} (-\hat y) =U_{c}+U_{0}\cos\left({k_x\frac{x-x_1}{L_x-x_1}}\right),
\label{vz_cosine}
\end{eqnarray}
}
Here, $U_{c}$ represents an offset, and $U_{0}$ is the strength of the 
shear variation of the ion flow. 
The mode number $k_x= {n\pi}/{2},~n=0.25 ,0.5,..1,$ represents 
the zeros of the corresponding monotonic ion velocity profile 
coinciding with the external 
boundary location $L_x$. The corresponding $\omega_{s}$ in the $xy$-plane
is found to be
{
\begin{eqnarray}
\omega_{s}=\nabla\times {\bf v}_i=-U_{0}\frac{k_x}{L_x-x_1} 
\sin\left({k_x\frac{x-x_1}{L_x-x_1}}\right),
\label{omega_s}
\end{eqnarray}
}

The entire analysis is done using 
the same driver velocity ${\bf v}_{i}(x,y)$, $n=1$, and the
corresponding vorticity $\omega_{s}(x,y)$. However, the similar 
vorticity $\omega_{s}(x,y)$ can be achieved for ions flow 
${\bf v}_{i} (-\hat x)$ with shear variation along 
the $\hat y$-direction.
For the $\omega_{B\|}$, the horizontal $B~\hat z$ is also considered having a monotonic shear profile over the bounded dust domain given by single natural mode of the cartesian plane as given by,
{
\begin{eqnarray}
{\bf B}\hat z=B_{0}\sin(yy),~~yy= k_y\frac{y-y_1}{L_y-y_1}.
\label{By_cosine}
\end{eqnarray}
}
The profile maintains minimum {\bf B} at the 
lower boundary $y=y_1$, $(y_1\approx 0)$ and a
maximum at the upper confining boundary $y=L_y$ of the bounded 
dust domain as highlighted in the Fig~\ref{schematic}. 
The mode number $k_y= {m\pi}/{2},~m=0.25, 0.5,.. 1$ represents 
the strength of the monotonic shear ${\bf B}$ that can varying 
through the $k_y$. In the Eq.~(\ref{E_a_eqn}) and Eq.~(\ref{cross_Ea_eqn5}), 
the diamagnetic drift contribution follows 
the similar form with that of magnetic-gradent drift when both the 
shear profile are natural mode of the cartesian plane. 
And our interest is mainly the analysis of 
${\bf B}$ effects on the dust vortex dynamics. Therefore, we ignored the diamagnetic drift for simplicity although any realistic plasma has the density gradient. 
Thus, Putting the Eq.~(\ref{By_cosine}) in Eq.~(\ref{cross_Ea_eqn4}), 
the simplified $\beta\omega_{B\|}$ in the ${\bf xy}$-plane
is found to be
{
\begin{eqnarray}\nonumber
\beta\omega_{B\|}=-\eta\frac{n_eq_eq_d}{m_d(1+\nu_{en}^2/\omega_{ec}^2)} 
\bigg[\bigg(\frac{-E_s}{B_{0}}\bigg)\bigg( \frac{k_y}{L_y-y_1}\bigg)\frac{cos (yy)}{sin^2(yy)}\\ 
+\bigg(\frac{K_bT_e}{q_eB_0}\bigg) \bigg(\frac{k_y}{L_y-y_1}\bigg)^2
\bigg(\frac{1}{sin(yy)}-\frac{2 cos^2(yy)}{sin^3(yy)} \bigg)
\bigg].
\label{omega_B_pll}
\end{eqnarray}
}
It is noted that $\omega_{B\|}$ consists of two main parts, i.e., the ${\bf E\times B}$-drift of both electrons and ions toward $-\hat x$-direction and the ${\bf \nabla B\times B}$-drift of electron toward the $\hat x$-direction while ions drift in the opposite direction. However, 
the shear nature of both the drifts acts together as the vorticity source. The last-term is three orders stronger than the first-term and six orders larger than the second-term which is almost negligible.   

Now, for defining the co-efficient $\beta$, $\xi$, $\nu$, 
and $\mu$, we consider a typical laboratory glow discharge argon 
plasma having  
dust density $n_d \simeq 10^{3}$ cm$^{-3}$,
ions density $n_i\simeq 10^{8}$ cm$^{-3}$,
neutral density $n_n\simeq 10^{15}$ cm$^{-3}$ (corresponds to 
$p\approx 12.4$ pascal of neutral pressure)~\cite{doi:10.1063/1.5078866},
electron temperature $T_{e}\simeq 3 eV$, 
ion temperature $T_{i}\simeq 1 eV$, and shear ions are streaming 
with $U_{0}$ cm/sec 
a fraction of the ion-acoustic 
velocity $c_{si}=\sqrt{T_{e}/m_{i}}\approx 10^{5}$ cm/sec 
while the dust acoustic velocity ($c_{sd}$) is $\approx 12~cm/sec$.
Here, $c_{ds}=\sqrt{Z_d^2(n_d/n_i)(T_i/m_d)}$, 
$Z_d\approx 10^4$ is dust charge number, 
$m_d\approx 10^{-14}~kg$ is dust mass, 
and other notations are all conventional~\cite{Merlino_2012}.
Using dust mass $m_d$, dust charge $q_d$, the width of the 
confined domain $ L_x(\approx 10$ cm), and steaming shear ions 
velocity strength $U_{0}$ as the ideal normalization 
units, the corresponding value of 
the system parameters/co-efficient are $\xi\approx 10^{-5}~ U_{0}/L_x$, 
$\nu\approx 10^{-2}~ U_{0}/L_x$, and 
$\mu\approx 2.5\times 10^{-4}~ U_{0}L_x$ 
respectively. We estimate the range of $\mu$ keeping 
in view that dust fluid flow is incompressible $u <c_{sd}$, and 
the associated Reynolds number is very small (Re $\approx 1$). 
This lead to a linear limit of the formulation
~\cite{PhysRevE.91.063110,doi:10.1063/1.4887003} and 
closely agree with that of Yukawa systems~\cite{doi:10.1063/1.1459708,PhysRevLett.88.065002}.  
Further, for a weak magnetized system of $B_0=4-500~G$ 
(corresponds to $\omega_{ec}\ge\nu_{en}$ to $\nu_{in}\ge\omega_{ic}$ ~\cite{doi:10.1063/1.4981928,doi:10.1063/1.4904039}), 
the normalized values of other system parameters are estimated as 
$B_0 \approx 1.6\times 10^{-7}[m_dU_0/q_dL_x]$ for $B_0=10~G$, the sheath electric field 
$E_s(=m_dg/q_d) \approx 9.8\times 10^{-5}[m_dU_0^2/q_dL_x]$ for $E_s\approx 61.25~V/m$, $n_e\simeq 9.0\times10^{7}$ cm$^{-3}$ satisfying the quasinutrality condition,
$(1+\nu_{en}^2/\omega_{ec}^2)\approx 1$, and
$\beta(= \eta n_eq_eq_d/m_d(1+\nu_{en}^2/\omega_{ec}^2))\approx 4.5\times 10^{-11}[U_0/L_x]$ 
respectively. In the calculation of resistivity $\eta$, we used 
standard Spitzer model with two order higher concerning the presence of dust and neutral particles~\cite{LI20162540,chen2015introduction}. 
\subsection{Dust flow characteristics driven 
by the $\omega_s$ without the magnetic 
field ($\omega_{B\|}=0$) ;
\label{Results_1}
}
The steady-state converged solutions are obtained for the bounded dust dynamics represented by the above set of equations~(\ref{streamf-equation}) and (\ref{vort-equation}) in the rectangular domain $0\le x\le 1$ and $0\le y\le 1$ near the sheath region as highlighted in Fig~\ref{schematic}. We adopt no-slip boundary conditions for all the physical boundaries confining the dust fluid~\cite{doi:10.1063/1.5045772}. 
\begin{figure}
    \includegraphics[width=.69\textwidth]{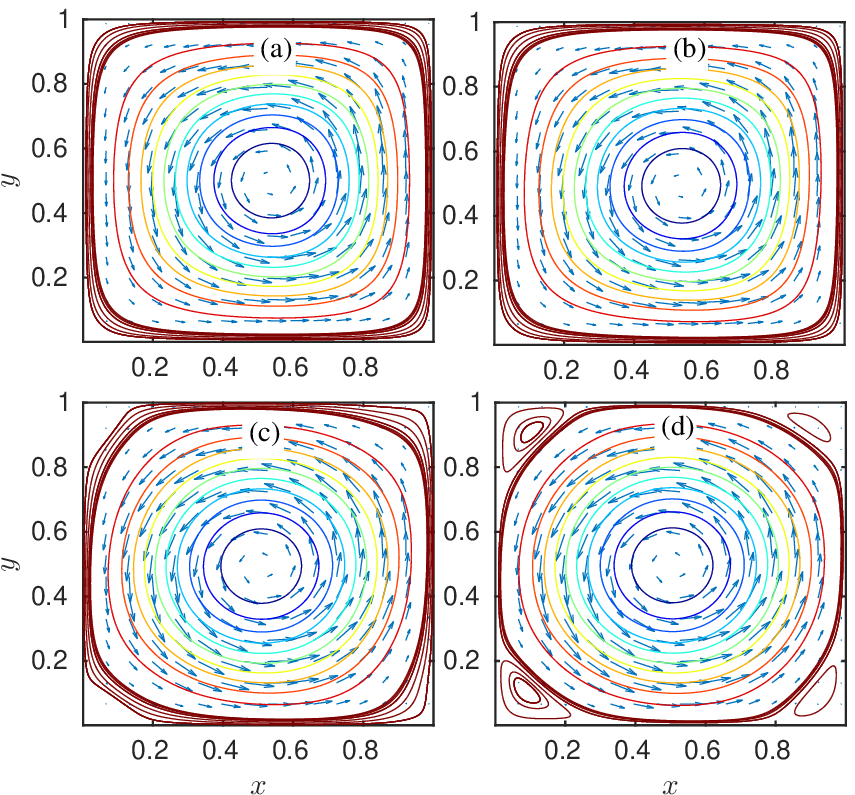}\\
     \includegraphics[width=.69\textwidth]{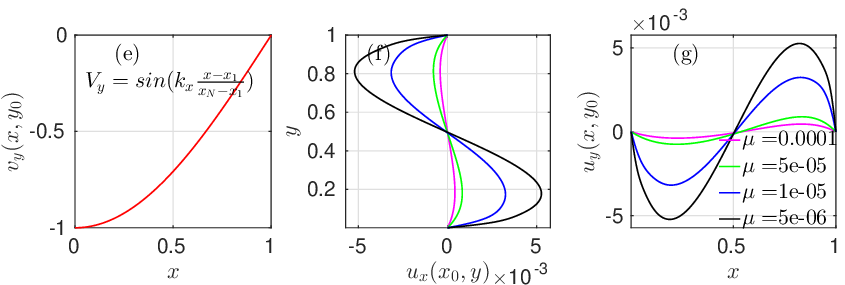}
 \caption{\small Streamlines for the steady bounded dust flow in the $x$-$y$ plane for varying 
 $(a)~\mu = 10^{-4} U_{0}L_{r}$, 
 $(b)~\mu = 5\times 10^{-5}U_{0}L_{r}$
 $(c)~\mu = 1\times 10^{-5}U_{0}L_{r}$
 and $(d)~\mu = 5\times 10^{-6} U_{0}L_{r} $ respectively having fixed other system parameters 
$\xi = 10^{-5} U_{0}/L_{r}$, 
$\nu =  10^{-4} U_{0}/L_{r}$, and 
${\bf B}=0$. 
(e) Cross-section profile of driver ion's velocity having sheared mode numbers $n=1$. The corresponding dust flow velocity profiles through the static point $(x_0,y_0)$ are (f) $u_x(x,y_0)$ and (g) $u_y(x_0,y)$ respectively. 
 \label{fig_diff_mu}
}
\end{figure}
The series of structural changes in terms of the streamlines and velocities profile of the bounded dust flow are presented in Fig~\ref{fig_diff_mu}, for a wide range of $\mu$ from $\mu = 10^{-4} U_{0}L_{x}$ to $\mu =5\times 10^{-6} U_{0}L_{x}$ and 
fixed other system parameters. Along with the structural change in Fig~\ref{fig_diff_mu}(a) to (d), the nonlinear structural bifurcation takes place through a critical $\mu^*$ and arise new structure having a circular core region surrounded by weak and elongated vortices near the corners. The corresponding changes in velocities strength and slight variation in boundary layer thickness are shown in Fig~\ref{fig_diff_mu}(f) and Fig~\ref{fig_diff_mu}(g). Boundary layers are the high shear region near the external no-slip boundaries~\cite{PhysRevE.95.033204}.
It demonstrates that the steady-state dust flow structure changes from linear to highly nonlinear regime with decreasing $\mu$, giving a new structure that retains more momentum and energy. In the previous work~\cite{PhysRevE.95.033204,Laishram_2019} without the magnetic field (${\bf \omega_{B\|}}=0$), similar solutions are analyzed in a 
cylindrical $rz$-plane using different parameters regimes, driving field, and boundary conditions. However, we reproduced it for comparison with the following new parametric effects in the present 
analysis in a cartesian setup.

\begin{figure}
    \includegraphics[width=.69\textwidth]{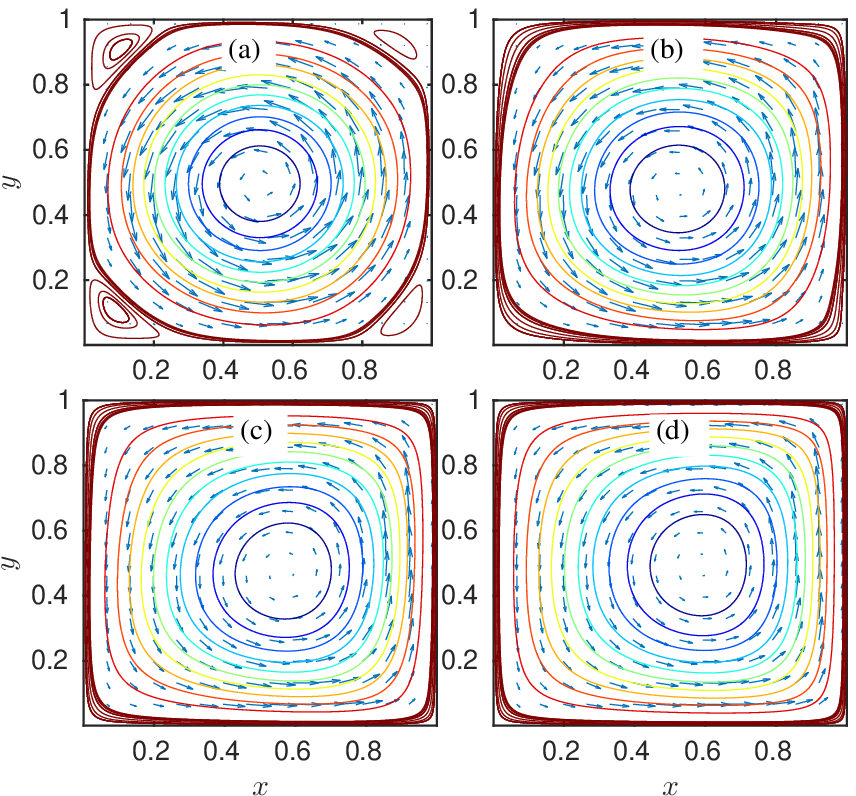}\\
    \includegraphics[width=.69\textwidth]{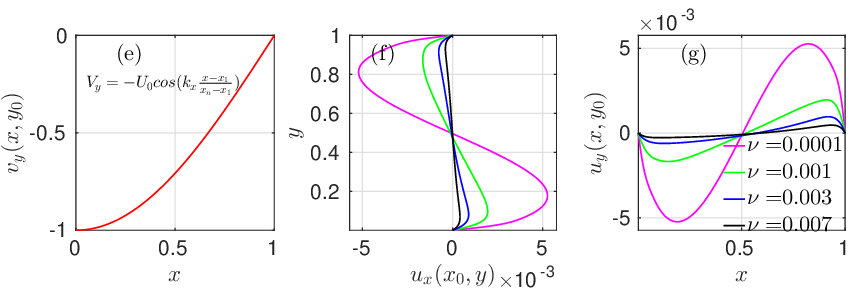}
  \caption{\small Streamlines for the steady bounded dust flow in the $x$-$y$ plane for varying 
 $(a)~\nu = 10^{-4} U_{0}/L_{r}$, 
 $(b)~\nu = 1\times 10^{-3}U_{0}/L_{r}$
 $(c)~\nu = 3\times 10^{-3}U_{0}/L_{r}$
 and $(d)~\nu = 7\times 10^{-3} U_{0}/L_{r} $ respectively having 
 fixed other system parameters 
$\xi = 10^{-5} U_{0}/L_{r}$, 
$\mu =  5\times 10^{-6} U_{0}L_{r}$, and 
${\bf B}=0$. 
(e) Cross-section profile of driver ion's velocity having sheared mode numbers $n=1$. The corresponding 
dust flow velocity profiles through the static point $(x_0,y_0)$ are (f) $u_x(x_0,y)$ and (g) $u_y(x,y_0)$ respectively. 
 \label{fig_diff_nu}
}
\end{figure}

Now, starting from the highly nonlinear flow structure shown in Fig~\ref{fig_diff_mu}(d), a series of steady-state dust flow structure in term of streamlines and velocity profile for a reasonable range of neutral collision frequency $\nu$ (corresponds to the pressure of $\approx 0.05-10~Pascal$ range) and fixed other system parameters is presented in Fig~\ref{fig_diff_nu}.  
In the series of structural change \ref{fig_diff_nu}(a) to (d), it is observed that neutral pressure takes a very sensitive role in determining the characteristic features of the bounded dust flow. The 
sensitivity of $\nu$ can be seen in the noticeable variation of boundary layer thickness in the Fig~\ref{fig_diff_nu}(f) and Fig~\ref{fig_diff_nu}(g).  The most probable reason appears in the model Eq.~(\ref{ns-equation}), in which the dust cloud is driven by the shear ion drag and the dissipative resistance is produced by both the viscous diffusion and collision with the background neutral fluid. Moreover, the neutral collision with ions can effects the momentum transfer from the ions to the dust and further reduces the strength of the dust circulation.

Thus, the following dust vortex characteristics are observed by comparing the series of flow structure in the Fig~\ref{fig_diff_mu} and Fig~\ref{fig_diff_nu}. First, in the weak flow regime (for higher $\mu$ 
and $\nu$), the forcing shear ions provide the rotary motion and its direction, while the confined domain determines its shape. The boundary layer is thick and its effects are distributed through the interior domain. Second, in the high flow regime (for lower $\mu$ and $\nu$), the nonlinear convective flow dominant over the diffusion, and the boundary effects are confined 
in thinner layers, so the interior flow responds to the monotonic forcing shear ion drag only. As a consequence, the vortices turn circular, and the rest of the domain is filled with several weaker vortical structures. Nevertheless, the size of secondary vortices increases as the nonlinearity increases and most of the circulating structure is 
co-rotating in nature following the shear scale of the driver ion's field.  Interestingly, Fig~\ref{fig_diff_nu}(f) and Fig~\ref{fig_diff_nu}(g) demonstrate that the ions drag-driven dust dynamics become negligibly small at higher $\nu\approx 7\times10^{-3}U_0/L_x$ (corresponds to neutral pressure of $\approx 10~Pascal$) even with the higher $\mu$ regime as reported in the laboratory experiment~\cite{doi:10.1063/1.5078866}.      
\begin{figure}
    \includegraphics[width=.69\textwidth]{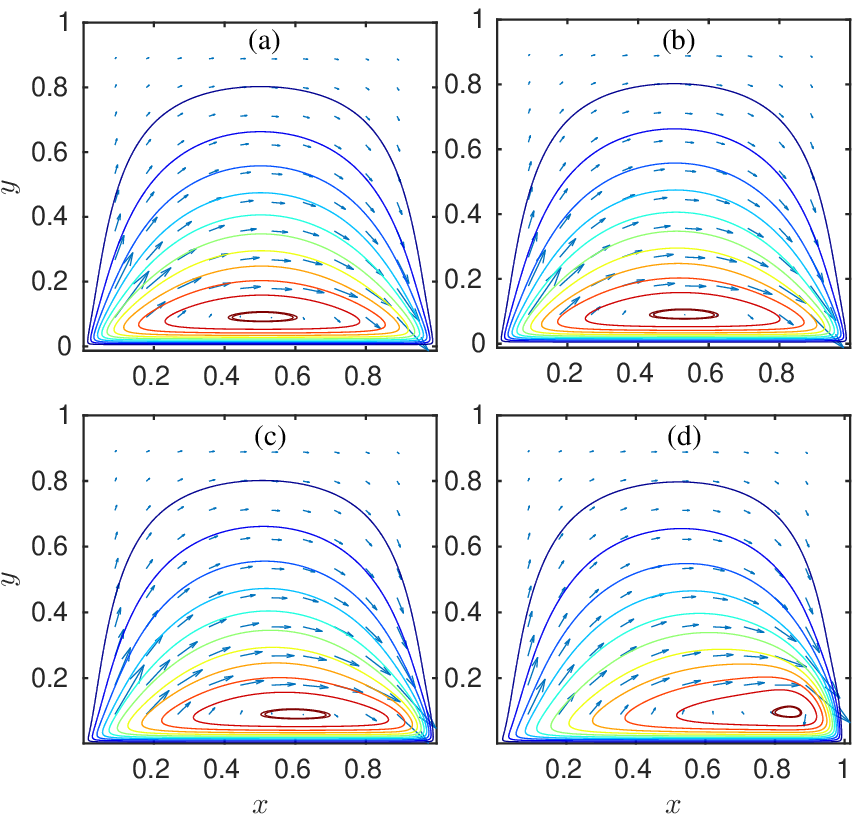}\\
   \includegraphics[width=.69\textwidth]{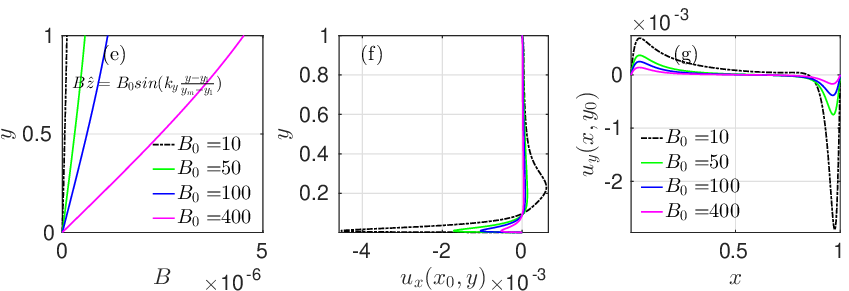}
\caption{\small Streamlines for the steady bounded dust flow in the $x$-$y$ plane for varying 
 $(a)~B_0 = 400~G$, 
 $(b)~B_0 = 100~G$
 $(c)~B_0 = 50~G$
 and $(d)~~B_0 = 10~G$ respectively having fixed other system parameters 
$\nu = 10^{-2} U_{0}/L_{r}$,  
$\mu = 5\times10^{-6} U_{0}L_{r}$, and 
$\omega_s=0$.
(e) The corresponding ${\bf B}$ profile for varying ${B_0}$ and fixed $k_y=\pi/4$. The dust flow velocity profiles through the interior static point $(x_0,y_0)$ are (f) $u_x(x_0,y)$ and (g) $u_y(x,y_0)$ respectively. 
 \label{fig_diff_B}
}
\end{figure}
\begin{figure}
      \includegraphics[width=.69\textwidth]{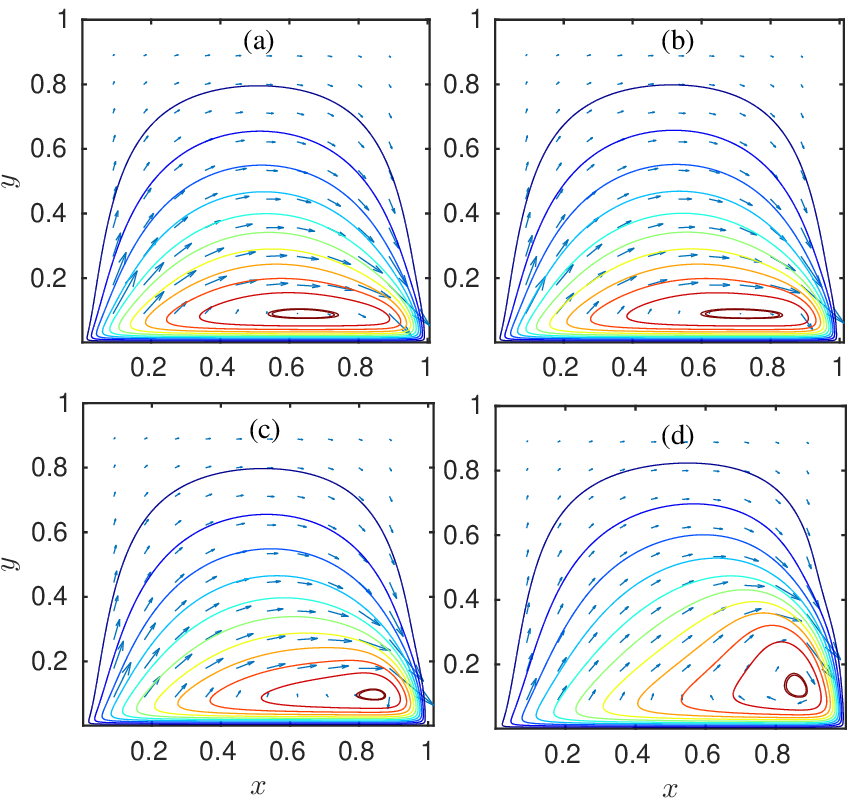}\\
    \includegraphics[width=.69\textwidth]{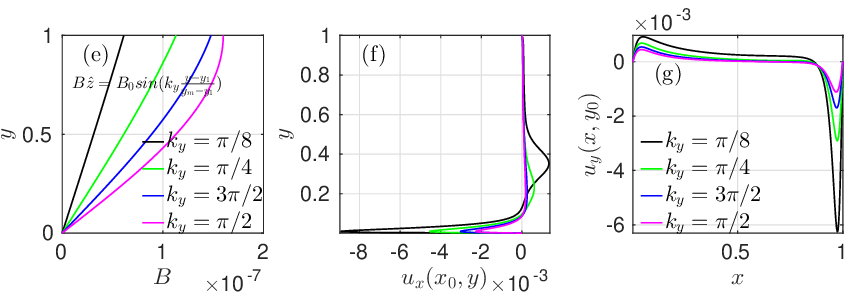}
 \caption{\small Streamlines for the steady bounded dust flow in the $x$-$y$ plane for varying 
 $(a)~k_y = \pi/2$, 
 $(b)~k_y = \pi/2$
 $(c)~k_y = \pi/2$
 and $(d)~~k_y = \pi/2$ respectively having fixed other system parameters 
$\nu = 10^{-2} U_{0}/L_{r}$, 
$\mu = 5\times10^{-6} U_{0}L_{r}$ and 
$\omega_s=0$.
(e) The corresponding ${\bf B}$ profile for varying $k_y$ and fixed ${B_0}=10~G$. The dust flow velocity profiles through the interior static point $(x_0,y_0)$ are (f) $u_x(x_0,y)$ and (g) $u_y(x,y_0)$ respectively. 
 \label{fig_diff_gradB}
}
\end{figure}
\subsection{Dust vortex characteristic driven 
by the $\omega_{B\|}$ in the weakly magnetized regime $~(\omega_{ic}\ll \nu_{in})$;
\label{structure_all_B}}
The analysis here is further extended in the high-pressure regime ($\nu=10^{-2}U_0/L_x$ where the ions drag-driven source $\omega_s$ is negligibly small and the driving field $\omega_{B\|}$ influences the dust dynamics. A series of structural changes in terms of the streamlines and velocities profile of the bounded dust flow are presented in Fig~\ref{fig_diff_B}, for the case $k_y=\pi/4$, in the wide range of magnetic field $B_0=10 G$ to $400 G$, and fixed other system parameters in the nonlinear convective regime of $\mu =  5\times10^{-6} U_{0}L_{x}$ and $\xi = 10^{-5} U_{0}/L_{x}$ respectively. It simply demonstrates that the $\omega_{B\|}$ due to the shear nature of $E_a$ can generate 
dust vortex dynamics.
Further, in the higher $B_0$ regime, the streamlines pattern in 
the Fig~\ref{fig_diff_B}(a) to Fig~\ref{fig_diff_B}(b) and the corresponding velocity profiles in Fig~\ref{fig_diff_B}(f)-(g) 
shows that the dust cloud circulates very slowly in a $D$-shaped elliptical structure. With decreases in $B_0$, the flow is strengthened gradually, the interior static point ($x_0, y_0$) convected toward the axial $\hat x$-direction, and the circulation turns into a small meridional structural as shown in Fig~\ref{fig_diff_B}(d). In the present analysis, the flow structure is 
determined by three main factors; the first one is the role of $E_s(\approx m_dg/q_d)$ in finding the static point $(x_0,y_0)$ which will be discussed in detail in the following section~\ref{structure_all_B_i}. The second is the profile of $\omega_{B\|}$ which is inversely proportional to $B_0$ ($sin(y)$), and the third is the incompressibility of the dust cloud. In the Eq.~(\ref{omega_B_pll}),
$\omega_{B\|}$ generates a localized but strong axial flow of the dust particles toward the $-\hat x$-direction. Because of the 
continuity, the localized flow is compensated by the whole clockwise circulation of the dust cloud in the $xy$-plane. Thus, the $D$-shaped elliptical structure is developed and it turns into a meridional structure with an increase in the nonlinear convective transport. The structural changes in the highly viscous linear regime $\mu \ge 1\times10^{-4} U_{0}L_{x}$ is indistinguishable in nature.

A more visible structural change is observed by varying the magnetic shear $k_y=m\pi/2$ instead of the field strength $B_0$. In the Fig~\ref{fig_diff_gradB}, a series of structural changes in terms of the streamlines and the velocities profile of the bounded dust flow are presented for $B_0=10~G$, a wide range of $ m=0.25$ to $1$, and other system parameters remain unchanged. The cross-section profile of the magnetic field for the varying $k_y$ is shown in Fig~\ref{fig_diff_gradB}(e). 
In comparison to Fig~\ref{fig_diff_B}, this analysis demonstrates that the dust peak velocities, the convection of ($x_0,y_0$), and meridional structural become more significant with a decrease in the $k_y$ than the decrease in $B_0$. Further, as in previous work of mode analysis~\cite{PhysRevE.91.063110}, the present analysis also allows us to examine the formation of multiple counter-rotating vortices by using the non-monotonic higher magnetic mode number $ m> 1$, says m=2, 3, or higher numbers in the Eq.~(\ref{By_cosine}). Interestingly, the $D$-shaped elliptical flow characteristics have various similarity, for instance, the $D$-shaped elliptical structure is common in MHD solution of 
magnetic confinement configurations (tokamak)~\cite{8692190,doi:10.1063/1.3613680} and in neutral flows in a lid-driven cavity at low to moderate Reynolds number~\cite{roy2020improved,articleMat}. In more specific, the meridional structure 
of dust circulation displayed in the Fig~\ref{fig_diff_gradB}(d) is observed in several magnetized~\cite{PhysRevLett.111.185003,doi:10.1063/5.0004842} and non-magnetized~\cite{PhysRevLett.101.235001} dusty plasma experiments although the parametric regimes and driving fields are different. It justifies the fact that these structures are among the common nonlinear characteristic features of driven-dissipative flow systems and dusty plasma provides a domain for analysis of the associated basic underlying physics.

\subsection{Dust vortex characteristic driven 
by both $\omega_s$ the $\omega_{B\|}$ ;
\label{structure_all_B_i}}
\begin{figure}
     \includegraphics[width=.69\textwidth]{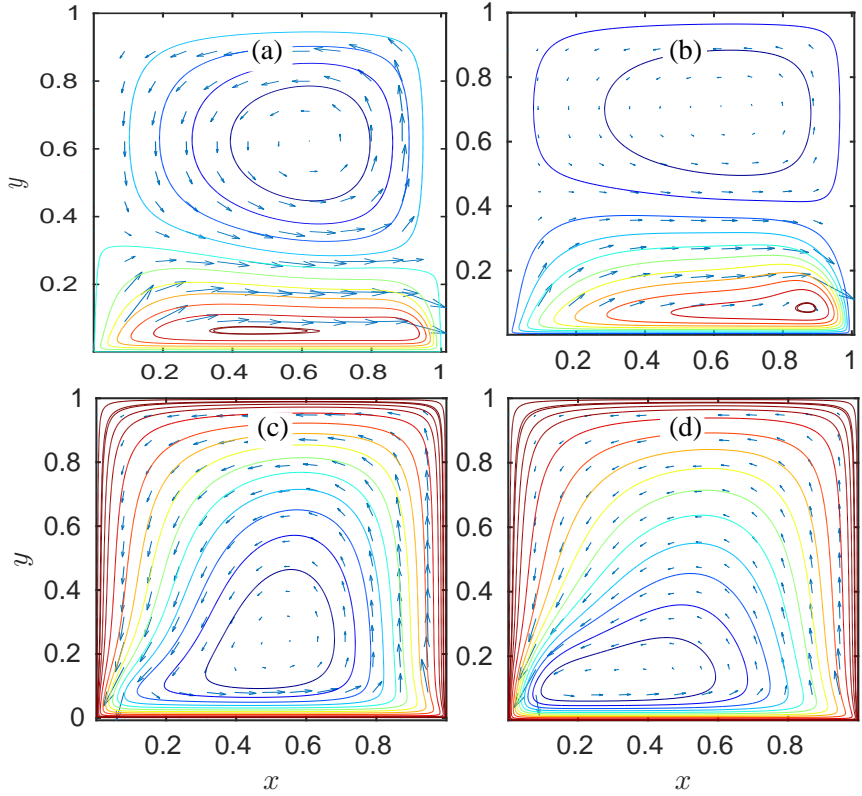}
 \caption{\small
($1^{st} ~row$) Streamlines for the steady bounded dust flow in the $x$-$y$ plane driven by both $\omega_s$ and $\omega_{B\|}$ for varying $(a)~k_y = \pi/2$ and $(b)~k_y = \pi/4$ respectively, having fixed other system parameters at
$\xi = 10^{-5} U_{0}/L_{r}$, 
$\nu = 10^{-2} U_{0}/L_{r}$, 
$\mu = 5\times10^{-6} U_{0}L_{r}$, and $B_0=10 G$. 
 ($2^{nd} ~row$) the corresponding streamlines pattern for varying $(c)~k_y = \pi/2$ and $(d)~k_y = \pi/4$ respectively, and negative $B_0=-10 ~G$.
 \label{fig_diff_modes}
}
\end{figure}

In this section, we finally examine the dust flow structure in 
the presence of both $\omega_s$ and the $\omega_{B\|}$ in the same configuration. We note from the above analysis in section~\ref{Results_1}, that the ion-drag produces the dust cloud circulation in an anti-clockwise direction and its strength reduces with increases in the neutral pressure. Again in section~\ref{structure_all_B}, weak magnetization generates dust cloud circulation in a clockwise direction and the strength increase with a decrease in $B_0$ and the $k_y$. Therefore, there are parametric regimes of high pressure and weak magnetization where both are comparable although 
ions-drag dynamics are dominant in most of the cases.
Now, we present the steady-state dust flow structure in the specific parametric regime of $\nu=10^{-2}U_0/L_x$, 
$\mu=5\times10^{-6}U_0L_x$, $\xi=10^{-5}U_0/L_x$, 
$B_0=10~G$, and $k_y=\pi/2$ where both $\omega_s$ and 
the $\omega_{B\|}$ are significant as shown 
in Fig.~\ref{fig_diff_modes}(a). 
\begin{figure}
     \includegraphics[width=.69\textwidth]{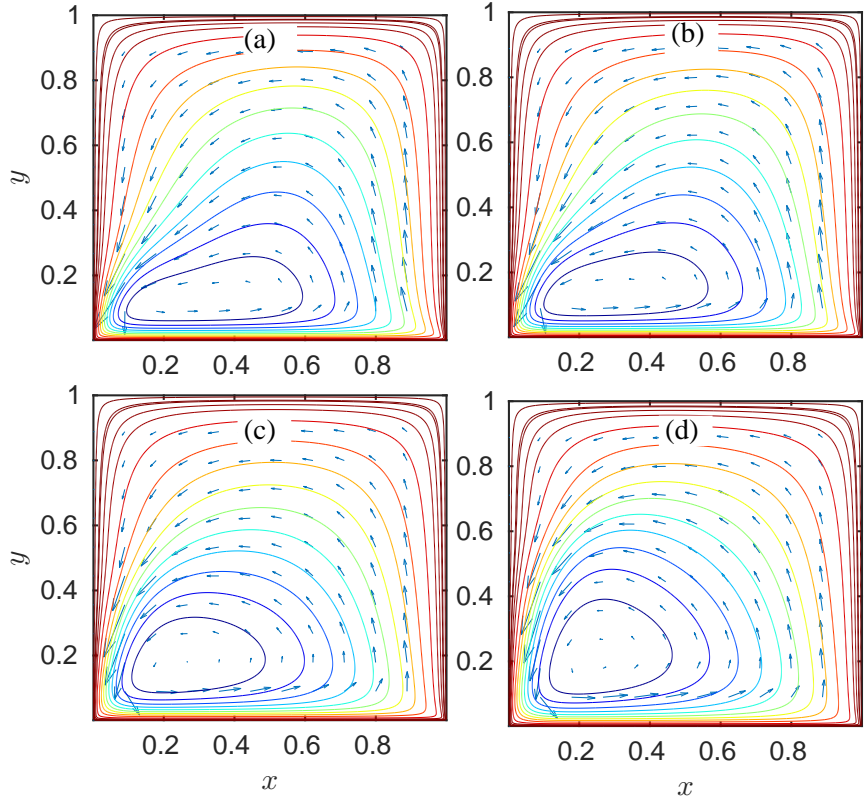}
 \caption{\small
 Streamlines for the steady bounded dust flow in the $x$-$y$ plane driven by both $\omega_s$ and $\omega_{B\|}$ for varying 
 $(a)~E_s = 61.25 ~V/m $, 
 $(b)~E_s = 306.25 ~V/m$
 $(c)~E_s = 612.5 ~V/m$
 and $(d)~~E_s = 3062.5 ~V/m$ respectively having fixed other system parameters 
$\xi = 10^{-5} U_{0}/L_{r}$, 
$\nu = 10^{-2} U_{0}/L_{r}$, 
$\mu = 5\times10^{-6} U_{0}L_{r}$, 
$B_0=-10~G$, and $k_y = \pi/4$.
 \label{fig_diff_E}
}
\end{figure}
%
It shows that unequal strength counter-rotating vortex pairs comprising of circular one driven by the ion-drags and localized elliptical structure driven by the weak magnetization can co-exist in the same plane. Further, in the case of lower $k_y=\pi/4$, the strength and size of the elliptical structure increase while the circular one is relatively reduced as shown in Fig.~\ref{fig_diff_modes}(b). While the parametric evolution of the unequal vortex pairs and its stability is deserving a separate analysis, the present work shows the feasibility of existing steady-state counter-rotating vortex pair of unequal strength associated with different monotonic shear flow fields as reported in the recent dusty plasma experiments~\cite{doi:10.1063/5.0004842}. Such unequal strength vortex-pair is common in vortex shedding dynamics behind aircraft wings~\cite{doi:10.1146/annurev-fluid-122414-034558}. 
Furthermore, when the direction of ${\bf B}$ is changed by 
$180^0$-degree, both the sources  $\omega_s$ and the $\omega_{B\|}$ act in the same direction and generate circular dominant meridional structural as shown in Fig.~\ref{fig_diff_modes}(c). It simply illustrates that the direction of magnetization is a key factor in the MDP dynamics. The corresponding circulation with lower $k_y=\pi/4$ is again plotted in Fig.~\ref{fig_diff_modes}(d), showing an elliptical dominant strong meridional structural due to the strengthening 
of $\omega_{B\|}$.  

One more interesting characteristics of the driven-dust vortex structure 
is the effect of varying ${\bf E_s}$. The cases with a varying value of 
${\bf E_s}$ is examined and the corresponding structural changes in terms of the streamlines are presented in Fig~\ref{fig_diff_E}, showing an increase in strength of the circulation and the interior static point ($x_0,y_0$) is convected toward the center of the ${\bf xy}$-plane. 
The role of the ${\bf E_s}$ can be addressed in two ways. In the first, $\omega_{B\|}$ is directly proportional to ${\bf E_s}$ in the Eq.~(\ref{omega_B_pll}), therefore, the strength of the circulation increase with increasing ${\bf E_s}$. In the second, applying the force balance condition~\cite{doi:10.1063/1.4904039} at the static equilibrium point ($x_0,y_0$) in Eq.~(\ref{ns-equation}), we have  

\begin{eqnarray}\nonumber
0 =-\nabla{\phi_b} +\frac{q_d}{m_d}{\bf E_a}-\frac{\nabla{P}}{\rho}  
+\mu\nabla^2{\bf u}+\xi{\bf v} +\nu{\bf w}.
\label{ns-equation_0}
\end{eqnarray}

Where, ${\bf E_s}(\approx m_dg/q_d)$ along $-\hat y$ is part of the confining potential ${\phi_b}$ againt the gravity. From the Eq.~(\ref{E_a_eqn}), the internal field ${\bf E_a}$ along $\hat x$ is directly dependent on the ${\bf E_s}$. Therefore, ${\bf E_s}$ takes an important role other than $\xi$, $\nu$, and $\mu$ in finding the steady-state dust flow structure in the confined domain. An increase in ${\bf E_s}$ means a rise in the levitation level of the dust cloud above the electrode.
The other parametric effects for $\xi$, $\nu$, and $\mu$ follow the same effect as discussed in the previous work~\cite{doi:10.1063/1.5045772,Laishram_2019}.  Most importantly, in all the case of the above analysis, the max dust speed is $u_d\le 6.0\times 10^{-3} U_{0}~i.e., ~u_d\le 6.0~cm/sec$ while the dust acoustic speed is $c_{ds}\approx~12.65~cm/sec$. This velocity range is in 
good agreement with several laboratory dust plasma experiments 
which have observed dust acoustic waves of speed in the range of 
12 cm/sec to 27.6 cm/sec~\cite{doi:10.1063/1.872238,
Merlino_2012}. 
\section{Summary and conclusions}
\label{conclusion}
The two dimensional(2D) hydrodynamic model for a bounded 
dust flow dynamics in plasma from the previous work~\cite{PhysRevE.95.033204} is extended 
for analysis of driven vortex characteristics in presence of external transverse and weak magnetic 
field (${\bf B}$) in a planner setup and parametric regimes motivated by recent magnetized dusty plasma (MDP) experiments.
We have demonstrated that the weak magnetization (only electron is magnetized) can produce an internal charge separation field ${\bf E_a}$ in between electrons and ions due to a combined effects of ${\bf E}\times {\bf B}$, ${\nabla n_e}\times {\bf B}$, ${\nabla {\bf B}}\times {\bf B}$-drifts, and collisions present in the system. Then the ${\bf E_a}$ can displace a dust particle or a dust cloud, whereas its shear can cause a rotation of the dust cloud. We derived the general equations for the ${\bf E_a}$ and associated 2D cross-section vorticity sources along with and across the ${\bf B}$ using the condition of 2D ambipolar diffusion in the dusty plasma.  
The formulation reveals that, in principle, vorticity can be developed on a plane parallel to the surface of electrodes in the setup, when there are variations of the ${\bf E_s}$ and magnetic/density shear along the ${\bf B}$. However, there is no contribution of the shear on the plane, therefore, difficult to achieve the condition in a real setup. This would be a probable reason why dust rotation was not observed in the real experiments by Puttscher and Melzer {\it et~al.}~\cite{doi:10.1063/1.4981928,doi:10.1063/1.4904039}.

Base on the proposed model, we have analyzed the characteristics of a 2D steady-state vortex developed on a plane perpendicular to the surface of the electrode and directed along the ${\bf B\hat z}$. The analysis is carried out assuming the ions flow profile of natural cosine mode, the magnetic field profile of natural cosine mode in the planner setup, and other parametric regimes are motivated by several recent MDP experiment. The analysis shows that the shear ions-drag force generates an anti-clockwise circular vortical structure and its characteristics are mainly depended on the kinematic viscosity and neutral collision frequency. The neutral pressure takes a key role in deciding the dominance between the ions-drag and the ${\bf E_a}$-force. When the neutral pressure is high, the ${\bf E_a}$-force becomes comparable or dominant over the ion-drag force and generates a strong localized flow of the dust particles. Then the localized flow is compensated by the whole 
incompressible dust cloud dynamics developing a clockwise $D$-shaped elliptical structure. 
Further, a decrease in ${\bf B}$, shows a noticeable increase of dust flow strength, convection of the interior static point, and the structure turns gradually into a meridional structure. Such elliptical and   
meridional structural are reported in several MDP experiments and various driven natural flow systems~\cite{doi:10.1063/5.0004842,PhysRevLett.111.185003,Nivedita@Dean_flow2017,doi:10.1146/annurev-fluid-122414-034558,marcus_1990}.

Additionally, we have examined the dust vortex dynamics in the parametric regimes of high pressure and low ${\bf B}$, where both the ions-drag and the magnetizing force are comparable although ions-drag dynamics is dominant in most of the cases. It has shown that unequal strength counter-rotating vortex pairs can co-exist in the same domain and their characteristics depend on the driving fields and the parametric regimes. When the direction of ${\bf B}$ is flipped by $180^0$-degree, both the drivers act together and give rise to a strong meridional structure. It simply points out the key role of ${\bf B}$-direction in MDP systems. 
Moreover, we analyzed the role of the ${\bf E_s}(\approx m_dg/q_d)$, which stands for the levitation level of the dust cloud above the electrode. It reveals that both the confinement potential and the ${\bf E_a}$ are directly proportional to 
the ${\bf E_s}$, therefore, any increase in ${\bf E_s}$ sharply enhanced the flow strength of the meridional structural and the convection of the interior static point toward the center of the plane. 

In conclusion, we stress that we have analyzed various mechanisms of dust vortex formation and its characteristics in a weakly magnetized dusty plasma. Specifically, we have interpreted the physics insight of elliptical vortex, meridional structural, and condition for co-existing unequal 
strength vortex pairs in several MDP, and also indicated the behaviors isomorphism with the relevant natural driven flows systems. This signifies the fact that dusty plasma can be an ideal domain for the study of various driven-dissipative flow systems. In the near future, a more comprehensive quantitative relation could be established between a 
threshold ${\bf B^*}$ and other parameters that overtake the ions drag force. And the parametric evolution of the vortex-pairs and its stability will be discussed using hydrodynamic stability analysis.\\
\section*{Acknowledgements}
Author L.~Modhuchandra acknowledges Dr. Devendra Sharma, Prof. Joydeep Ghosh, and Prof. Abijit Sen, Institute for Plasma Research, India, for the invaluable support 
and encouragement all the time. The numerical calculations in this paper used the resources of $Antya-cluster$ of Institute for Plasma Research, India.


%
\end{document}